\documentclass[11pt]{article}
\usepackage{latexsym}  
\usepackage{amssymb}
\usepackage{graphicx}
\usepackage{amsmath}

\begin{document}

\begin{center}
{\Large\bf O(3) Flavor Symmetry and }\\
{\Large\bf  an Empirical Neutrino Mass Matrix}

\vspace{3mm}
{\bf Yoshio Koide}

{\it IHERP, Osaka University, 1-16 Machikaneyama, 
Toyonaka, Osaka 560-0043, Japan} \\
{\it E-mail address: koide@het.phys.sci.osaka-u.ac.jp}

\date{\today}
\end{center}

\vspace{3mm}
\begin{abstract}
Based on a new approach to quark and lepton masses, where 
the mass spectra originate in vacuum expectation values
of O(3)-flavor {\bf 1}+{\bf 5} (gauge singlet) scalars, a neutrino
mass matrix of a new type is speculated.  The mass
matrix is described in terms of the up-quark 
and charged lepton masses, and, by assuming a special flavor basis, 
it can be accommodated to a nearly tribimaximal mixing without 
explicitly assuming a discrete symmetry.
Quark mass relations are also discussed based on the
new approach. 
\end{abstract}

\vspace{3mm}

{\large\bf 1 \ Introduction}
 
One of the most challenging problems in contemporary 
particle physics is to clarify the origin of flavors.
For this purpose, searching for a unified description
of the observed quark and lepton mass spectra will
provide a promising clue to us.
In conventional mass matrix model, the quark and lepton
mass matrices $M_f$ are given by the forms
$(M_f)_{ij} =(Y_f)_{ij} v_H$, where $(Y_f)_{ij}$ are
coupling constants of the Yukawa interactions
$\bar{f}_{Li} f_{Rj} H^0$ and $v_H$ is a
vacuum expectation value (VEV) of the neutral component
of the Higgs scalar $H$, $v_H =\langle H^0\rangle$.
Against this conventional approach, 
there is another idea: the origin of the mass spectra
is due to VEV structures of Higgs scalars $H_{ij}$
\cite{nonetHiggs,nonet-Higgs07}, 
i.e. $(M_f)_{ij} = y_f \langle (H^0)_{ij}\rangle$.
In the present paper, we will investigate an extended 
model by separating the role of $H_{ij}$ into two roles:
one of the roles is to cause SU(2)$_L$ symmetry breaking 
at the energy scale $\mu \sim 10^2$ GeV, and 
the conventional SU(2)$_L$ doublet Higgs scalars $H_u$ 
and $H_d$ still play the role in this scenario;
another one is to give an origin of the
mass spectra, and we consider gauge-singlet scalars
$(Y_f)_{ij}$ whose VEVs give effective Yukawa coupling
constants $\langle (Y_f)_{ij}\rangle/\Lambda$ 
($\Lambda$ is an energy scale of the effective theory).
As a typical model with such gauge-singlet scalars $(Y_f)_{ij}$,
there has been a model \cite{nonetVEV08} with U(3)-nonet
scalars, where quarks and leptons are assigned to 
{\bf 3} and $\bar{{\bf 3}}$ of U(3)$_F$, so that
$Y_f$ ($f=u,d,\nu,e$) are assigned to nonet of U(3)$_F$.
However, this U(3) scenario with this assignment cannot 
apply to a grand unification theory (GUT) scenario, 
because, for example, in SU(5)-GUT, the SU(2)$_L$ doublet
and singlet quark fields $Q$ and $U$ should be 
assigned to the same multiplet {\bf 3}, so that $Y_u$ must be 
$\bar{\bf 6}$ (not nonet).
Then, the model considerably become complicated,
because we need fields {\bf 6} in addition to fields
$\bar{\bf 6}$ in order to make singlets of U(3).

In this paper, considering applicability of the scenario 
to a GUT scenario, 
we assume an O(3) flavor symmetry instead of U(3). 
We consider the following superpotential terms:
$$
W_{Y}= \sum_{i,j} \frac{y_u}{\Lambda} U_i(Y_u)_{ij} {Q}_{j} H_u  
+\sum_{i,j}\frac{y_d}{\Lambda} D_i(Y_d)_{ij} {Q}_{j} H_d 
$$
$$
+\sum_{i,j} \frac{y_\nu}{\Lambda} L_i(Y_\nu)_{ij} {N}_{j} H_u  
+\sum_{i,j}\frac{y_e}{\Lambda} L_i(Y_e)_{ij} E_j H_d  
+h.c. + \sum_{i,j}y_R N_i (Y_R)_{ij} N_j ,
\eqno(1.1)
$$ 
where $Y_f$  ($f=u,d,\nu,e$) and $Y_R$ are O(3)-flavor 
{\bf 1}+{\bf 5} (gauge singlet) fields , 
and $Q$ and $L$ are quark and lepton 
SU(2)$_L$ doublet fields of O(3)$_F$ triplets,  
and $U$, $D$, $N$, and $E$ are 
SU(2)$_L$ singlet matter fields of O(3)$_F$ triplets.
Therefore, the fields $Y_f$ and ${Y}_R$ are symmetric.
Our basic assumption is as follows: the fields $Y_f$ and
$Y_R$ always behave as a combination of {\bf 1}+{\bf 5},
so that, for example, {\bf 5} alone never appears
in the interaction terms.
Hereafter, for convenience, we will denote {\bf 1}+{\bf 5} 
as {\bf 6}.  
(Note that if $Y_f$ are composed of a single ({\bf 1}+{\bf 5}),
$Y_f$ are real, but if $Y_f$ are composed of ({\bf 1}+{\bf 5})'s
more than two, $Y_f$ can be complex with $Y_f^T=Y_f$.)
In order to distinguish the fields $Y_f$ from each other, 
we assign additional U(1) charges $Q_X(Y_f)=q_f$ to $Y_f$
($f=u,d,\nu,e$),  
and $Q_X(U)=-q_u$ to $U$, $Q_X(E)=-q_e$ to $E$, and so on.
The field $Y_R$ has the charge $Q_X(Y_R)=2q_\nu$. 
In this paper, we will write down our superpotential $W$ 
under the O(3)$_F$ and U(1)$_X$ symmetries.

In the present approach, we will investigate relations 
among $Y_f$ and $Y_R$
by using supersymmetric (SUSY) vacuum conditions for
the superpotential $W=W_u+W_d+W_\nu +W_e+W_R+W_Y$, 
where $W_f$ ($f=u,d,\nu,e$) and $W_R$ determine the VEV
structures of $Y_f$ and $Y_R$, respectively. 
(Since we can easily show $\langle Q\rangle =\langle L\rangle 
=\langle U\rangle =\langle D\rangle =\langle N\rangle 
=\langle E\rangle =0$, hereafter, we will
drop the term $W_Y$ from $W$ when we investigate the VEV 
structures of $Y_f$. )
Such an approach to quark and lepton mass matrices has first been
adopted by Ma \cite{Ma-PLB07} and has been developed by the author 
within a context of U(3)-flavor nonet model \cite{nonetVEV08}.
In the conventional mass matrix approach, the investigation has 
now been on a level with theoretically reliable ground via a 
long period of phenomenological investigations. 
However, the present approach is still in its beginning
stage, so that  
we need more phenomenological investigations.
Therefore,  we adopt the following strategy in this
approach: 
(i) First, we search for a possible form of the superpotential
$W$ which can successfully provide relations among the 
observed masses and mixings from the phenomenological point
of view;
(ii) Next, we investigate what symmetries or quantum number
assignments can explain such a specific form of $W$.
In this paper, we will investigate a possible 
form of $W$ by putting weight on the step (i). 

Recently, as a byproduct in such approach,  
an interesting neutrino mass matrix form \cite{empiricalMnu08}
 has been reported: the form is given by  
$$
M_\nu \propto Y_e^{-1} Y_u^{1/2} + Y_u^{1/2} Y_e^{-1}+\xi_0 {\bf 1}.
\eqno(1.2)
$$
Neutrino mass matrix models which leads to the so-called 
tribimaximal neutrino mixing \cite{tribi} have usually been 
proposed based on discrete symmetries, 
while, if we assume a specific relation between a diagonal
basis of the charged lepton mass matrix (we call it ``$e$-basis")
and a diagonal basis of the down-quark mass matrix (we call it
``$d$-basis"), the mass matrix (1.2)
can be accommodate to a nearly tribimaximal neutrino mixing 
without explicitly assuming a discrete symmetry.
On the other hand, in general, if a neutrino mass matrix $M_\nu$ 
can give reasonable masses and mixing, 
a neutrino mass matrix $\tilde{M}_\nu$ with 
an inverse form of $M_\nu$, $\tilde{M}_\nu=m_0^2 M_\nu^{-1}$, 
can also give 
reasonable predictions, because, by taking the inverse of
$U^\dagger M_\nu U^* = M_\nu^D \equiv {\rm diag}(m_{\nu1}, m_{\nu2},
m_{\nu3})$, we can obtain $U^T \tilde{M}_\nu U =m_0^2 (M_\nu^D)^{-1}
={\rm  diag}(m_0^2/m_{\nu1}, m_0^2/m_{\nu2},  m_0^2/m_{\nu3})$,
i.e. we obtain the mixing matrix $U^*$ instead of $U$ and 
neutrino masses  
$(m_0^2/m_{\nu1}, m_0^2/m_{\nu2},  m_0^2/m_{\nu3})$
with a normal (inverse) hierarchy
instead of neutrino masses 
$(m_{\nu1}, m_{\nu2},m_{\nu3})$
with an inverse (normal) hierarchy.
Therefore, in this paper, instead of the model 
$M_\nu \propto Y_e^{-1} Y_u^{1/2} + Y_u^{1/2} Y_e^{-1} +\xi_0 {\bf 1}
=Y_e^{-1}(Y_u^{1/2} Y_e +Y_e Y_u^{1/2} + \xi_0 Y_e Y_e )Y_e^{-1}$, 
we will investigate a neutrino mass matrix with a seesaw-type
$$
M_\nu = \frac{y_\nu^2 v_{Hu}^2}{y_R \Lambda^2} Y_\nu 
Y_R^{-1} Y_\nu^T ,
\eqno(1.3)
$$
where $Y_R$ and $Y_\nu$ are given by
$$
Y_R \propto Y_u^{1/2} Y_e +Y_e Y_u^{1/2} + \xi_0 Y_e Y_e ,
\eqno(1.4)
$$
and $Y_\nu \propto Y_e$, respectively.
In the model (1.2), the matrix 
$M_\nu$ was for Dirac neutrinos, while the present 
$M_\nu$ is for Majorana neutrinos.  
The mass matrix (1.2) could not provide
a reasonable mass spectrum without adjusting the parameter
$\xi_0$,  
while, in this paper, we will give a small value of
$\Delta m_{21}^2/|\Delta m_{32}^2|$ 
without the $\xi_0$-term.
Note that, in the present scenario, since the Dirac neutrino 
mass matrix $Y_\nu$ is identical with the charged lepton mass
matrix $Y_e$, the nearly tribimaximal mixing originates in 
the structure of $Y_R$.

In the next section, we will  derive the neutrino mass 
matrix (1.3) with the form (1.4) of $Y_R$ by using
SUSY vacuum conditions for an O(3)$_F$ and U(1)$_X$ 
invariant superpotential, and we will evaluate the mass
matrix $M_\nu$ by using the observed values of up-quark and 
charged lepton masses.
However, in order to evaluate the neutrino mixing matrix,
we must know the form of (1.3) on the $e$-basis, especially 
the form of $Y_u^{1/2}$ on the $e$-basis.
Therefore, in the present paper, we will put a phenomenological
assumption on the relation between $e$- and $d$-bases.
Then, we will find that, by using the observed Cabibbo-Kobayashi-Maskawa 
(CKM) matrix parameters, the mass matrix (1.3) with the form (1.4)
can be accommodated to the observed nearly tribimaximal mixing.
However, since the result is dependent on a phenomenological
assumption on the form of $\langle Y_u\rangle$ on the basis ``$e$-basis",
the mass matrix is still an empirical one.  
Nevertheless, we consider that the result is very 
suggestive.

In Sec.3, we will discuss the structure of $Y_d$ lightly.
Finally, Sec.4 will be devoted to concluding remarks.

\vspace{3mm}

{\large\bf 2 \ Neutrino mass matrix without a discrete symmetry}

In order to give the operator $Y_u^{1/2}$
in the expression (1.4), we introduce additional O(3)$_F$ {\bf 6}
fields $\Phi_u$ and $X_u$ with the U(1)$_X$ charges 
$\frac{1}{2}q_u$ and $-q_u$, respectively.
Then, we can write down the superpotential for the $u$-sector
$$
W_u = \lambda_u {\rm Tr}[\Phi_u \Phi_u X_{u}]
+m_u {\rm Tr}[Y_u X_{u}] + W_{\Phi u}(\Phi_u) .
\eqno(2.1)
$$
From SUSY vacuum conditions (for the moment, we regard $W_u$
as $W$), we obtain
$$
\frac{\partial W}{\partial X_{u}} = 0
=\lambda_u \Phi_u \Phi_u + m_u Y_u ,
\eqno(2.2)
$$
$$
\frac{\partial W}{\partial Y_u} = 0
= m_u X_{u} ,
\eqno(2.3)
$$
$$
\frac{\partial W}{\partial \Phi_{u}} = 0
= \lambda_u (\Phi_u X_{u} + X_{u} \Phi_{u})
+ \frac{\partial W_{\Phi u}}{\partial \Phi_{u}}.
\eqno(2.4)
$$
From the condition (2.2), we obtain a bilinear relation
$$
\langle Y_u\rangle = - \frac{\lambda_u}{m_u} 
\langle\Phi_u\rangle \langle\Phi_u\rangle ,
\eqno(2.5)
$$
so that the field $\Phi_u$ plays a role of $Y_u^{1/2}$.
However, since the matrix $\langle \Phi_u \rangle$ is
not Hermitian, the relation
$$
U_u^T \langle Y_u \rangle U_u =
\langle Y_u \rangle^D \propto {\rm diag} ({m_{u1}},
{m_{u2}},{m_{u3}}) ,
\eqno(2.6)
$$ 
does not always mean
$$
U_u^T \langle \Phi_u \rangle U_u =
\langle \Phi_u \rangle^D \propto {\rm diag} (\sqrt{m_{u1}},
 \sqrt{m_{u2}},\sqrt{m_{u3}}) ,
\eqno(2.7)
$$ 
where $D$ denotes that the matrix is  on its diagonal basis.
As we see later, we need the relation (2.7).
Therefore, we assume the field $\Phi_u$ (and also $Y_f$)
is real, so that the matrix $U_u$ is orthogonal matrix.

From the condition (2.3), we obtain
$$
\langle X_{u}\rangle = 0 .
\eqno(2.8)
$$
Therefore, from the condition (2.4), we obtain
${\partial W_{\Phi u}}/{\partial \Phi_{u}} =0$.
We assume that 
three eigenvalues of $\langle \Phi_u \rangle$
can completely be determined by this condition
${\partial W_{\Phi u}}/{\partial \Phi_{u}} =0$.
However, for this purpose, the superpotential term 
$W_{\Phi u}$ will include U(1)$_X$ symmetry breaking terms.
In this paper, we do not discuss the explicit form
of $W_{\Phi u}$.
We assume that the VEV values are suitably given
by Eq.(2.7) with the observed up-quark masses 
$m_{ui}$.

For convenience, for the $e$-sector, we also assume 
superpotential term $W_e$ similar to the $u$-sector:
$$
W_e = \lambda_e {\rm Tr}[\Phi_e \Phi_e X_{e}]
+m_e {\rm Tr}[Y_e X_{e}] + W_{\Phi e}(\Phi_e) ,
\eqno(2.9)
$$
where $\Phi_e$, $X_{e}$ and $Y_e$ have U(1)$_X$ charges
$\frac{1}{2}q_e$, $-q_e$ and $q_e$, respectively, 
so that we obtain relations
$$
Y_e = - \frac{\lambda_e}{m_e} \Phi_e\Phi_e ,
\eqno(2.10)
$$
with $\Phi_e^D \propto {\rm diag} (\sqrt{m_{e1}},
 \sqrt{m_{e2}},\sqrt{m_{e3}})$,
where we have again assumed that the field $\Phi_e$ is real.
(Hereafter, for simplicity,  we will sometimes express 
VEV matrices $\langle A\rangle$ as simply $A$.)

In order to obtain the relation $Y_\nu \propto Y_e$,
we assume the following structure of $W_\nu$:
$$
W_\nu = \lambda_{\nu } \phi_\nu {\rm Tr}[Y_\nu X_\nu]
+\lambda_{\nu e} \phi_e {\rm Tr}[Y_e X_\nu],
\eqno(2.11)
$$
where $\phi_\nu$ and $\phi_e$ are gauge- and
flavor-singlet fields, and  we assign U(1)$_X$ charges
as $Q_X(X_\nu)=x_\nu$, $Q_X(\phi_\nu)=-(q_\nu+x_\nu)$ and
$Q_X(\phi_e)=-(q_e+x_\nu)$.
From $\partial W/\partial \phi_\nu=0$ and 
$\partial W/\partial \phi_e=0$, we obtain 
$X_\nu =0$.
From $\partial W/\partial X_\nu=0$, we obtain
$$
Y_\nu = -\frac{\lambda_{\nu e} \phi_e}{\lambda_\nu \phi_\nu}
Y_e .
\eqno(2.12)
$$

Next, let us  investigate a possible form of $W_R$.
In order to obtain the relation (1.4) from the phenomenological point
of view, we assume the following form of $W_R$:
$$
W_R = \lambda_R  {\rm Tr}[(Y_e \Phi_u +\Phi_u Y_e) X_R] 
+m_R {\rm Tr}[Y_R X_R] ,
\eqno(2.13)
$$
where we have assumed U(1)$_X$ charges $Q_X(Y_R)=-Q_X(X_R)=2 q_\nu$
and $\frac{1}{2}q_u +q_e -2q_\nu=0$,  
From SUSY vacuum conditions 
${\partial W}/{\partial {Y}_{R} } = 0$,  we obtain $X_R=0$.
Then, the requirement ${\partial W}/{\partial Y_e} = 0$ 
leads to the condition
${\partial W_e}/{\partial Y_e}=0$, so that we obtain
the relation (2.10).
From ${\partial W}/{\partial {X}_R } = 0$, we obtain
$$
Y_R = - \frac{\lambda_R}{m_R} ( Y_e \Phi_u + \Phi_u Y_e) .
\eqno(2.14)
$$
Thus, we can obtain the desirable form (1.4) of $Y_R$ 
(without the $\xi_0$-term).

For convenience, let us define a name of a flavor basis 
as follows:
when a VEV matrix $\langle Y_f\rangle$ takes a diagonal form 
on a basis,
we call the basis ``$f$-basis", and we denote a form 
of a matrix $\langle A\rangle$ on the $f$-basis as 
$\langle A\rangle_f$.
In order to obtain the neutrino mixing matrix form 
on the $e$-basis, 
we must know a matrix form of $\langle\Phi_u\rangle$ on 
the $e$-basis, i.e. $\langle\Phi_u\rangle_e$,
although the form $\langle\Phi_u\rangle_u$ on the $u$-basis is given
by Eq.(2.7).
Let us defined a transformation of a VEV matrix 
$\langle Y_f\rangle$ 
from  a $b$-basis to an $a$-basis as
$$
\langle Y_f\rangle_a = U_{ba}^T \langle Y_f\rangle_b U_{ba} ,
\eqno(2.15) 
$$
where $U_{ab}$ are unitary matrices, and they satisfy
$U_{ab}^\dagger =U_{ba}$ and $U_{ab}U_{bc}=U_{ac}$.
(These operators $U_{ab}$ are not always members of
O(3) flavor transformations.)  
Since $Y_f^T=Y_f$ in the present model, the VEV matrix 
$\langle Y_f\rangle$ are diagonalized as
$U_f^T \langle Y_f\rangle U_f = \langle Y_f\rangle^D$.
Therefore, $\langle Y_u\rangle_d$ is given by
$\langle Y_u\rangle_d =V^T(\delta) \langle Y_u\rangle_u
V(\delta)$, where $V(\delta)$ is the standard expression of 
CKM matrix.
The simplest assumption is to consider that the $d$-basis 
is identical with the $e$-basis, so that 
we can regard $U_{ue}$ as $U_{ue} =V$ because $U_{ud}= V$.
Then, we can evaluate the neutrino mass matrix (1.3) with
$\langle Y_R\rangle_e \propto \langle\Phi_u\rangle_e 
\langle Y_e\rangle_e + \langle Y_e\rangle_e  
\langle\Phi_u\rangle_e$ by using the form 
$$
\langle\Phi_u\rangle_e =U_{ue}^T \langle\Phi_u\rangle_u U_{ue} =
V^T(\delta) \langle\Phi_u\rangle^D V(\delta) .
\eqno(2.16)
$$
(Note that the O(3)-invariant relation (2.5) is not valid 
on the $d$-basis, because $U_{ud}=V(\delta)$ is not orthogonal,
although we can still use $\langle\Phi_u\rangle_d =
V^T(\delta) \langle\Phi_u\rangle_u V(\delta)$.
The relation (2.5) is valid only on a basis which is transformed 
from the $u$-basis by an orthogonal transformation.)
In the numerical calculation of $M_\nu$, we adopt 
the standard phase convention $V(\delta)$ \cite{PDG06}  
of the CKM matrix, and use the following
input values: the up-quark masses \cite{FK-qmass} at the 
energy scale $\mu=M_Z$,  
$m_{u1}=0.00127$ GeV, $m_{u2}=0.619$ GeV, $m_{u3}=171.7$ GeV,
 and the CKM parameters \cite{PDG06},
$|V_{us}|=0.2257$, $|V_{cb}|=0.0416$, $|V_{ub}|=0.00431$.
(Here, we have used the quark mass values at $\mu=M_Z$ because
we have used the CKM parameter values at $\mu=M_Z$.  For the
energy scale dependency of the mass ratios and CKM parameters,
for example, see Ref.\cite{evol}.)
As seen in Table 1, the results are dependent on the $CP$ 
violating phase parameter $\delta$.
The present experimental data \cite{PDG06} on the CKM
matrix favor $\delta \simeq \pi/3$.   
However, as seen in Table 1, the predicted value of 
$\sin^2 2\theta_{23}$ at  $\delta \simeq \pi/3$
is in poor agreement with the observed value 
 $\sin^2 2\theta_{23}=1.00_{-0.13}$ \cite{MINOS},
 although the predicted value of $\tan^2\theta_{12}$ is
 roughly in agreement with the observed value 
$\tan^2\theta_{12}=0.47^{+0.06}_{-0.05}$ \cite{KamLAND}. 
Therefore, we cannot regard that the $e$-basis is identical with
the $d$-basis.

\begin{table}
\caption{
The $\delta$ dependency of predicted values in the case
 $U_{ue}=V(\delta)$.
The values of $\sin^2 2\theta_{23}$ and $\tan^2 \theta_{12}$ are 
estimated by 
$\sin^2 2\theta_{23} = 4 |(U_\nu)_{23}|^2 |(U_\nu)_{33}|^2$ and
$\tan^2 \theta_{12}=|(U_\nu)_{12}/(U_\nu)_{11}|^2$, respectively.
The numerical results in the case $U_{ue}=V(-\delta)$ are identical
with the case $U_{ue}=V(\delta)$. 
}

\begin{center}
\begin{tabular}{|c|l|l|l|l|l|}\hline
$\delta$ &  $\sin^2 2\theta_{23}$ & $\tan^2 \theta_{12}$ &
$|U_{13}|$ & $\Delta m^2_{21}/\Delta m^2_{32}$ \\ \hline
0    &  $0.3890$ & $0.4679$ & $0.01156$ & $0.00220$ \\
$60^\circ$ & $0.7702$ & $0.4979$ & $0.01779$ & $0.00100$ \\
$90^\circ$ & $0.9237$ & $0.5228 $ & $0.01529$ & $0.00070$ \\
$120^\circ$ & $0.9836$ & $0.5434$ & $0.01055$ & $0.00063$ \\
$180^\circ$ &  $0.9998$ & $0.5604$ & $0.00034$ & $0.00062$ 
\\ \hline
\end{tabular}\end{center} 
\end{table}

However, as seen in Table 1, note that the case with 
$\delta \geq 2\pi/3$ can give a nearly
tribimaximal mixing.
Especially, we notice that the case $\delta=\pi$ highly
realizes the tribimaximal mixing.
Since we have assumed that $\Phi_u$ and $\Phi_e$ (therefore, 
$Y_u$ and $Y_e$) are real matrices, so that the $u$- and 
$e$-bases are connected each other by an orthogonal
transformation. 
This guarantees to use the relations (2.5), (2.12) and
(2.14), which are obtained from the O(3)-invariant superpotential,
on the $u$- and $e$-bases.
If we still suppose $U_{ue}\simeq U_{ud}=V(\delta)$,
the possible candidates of the
orthogonal matrix $U_{ue}$ will be $U_{ue}=V(0)$ and/or
$U_{ue}=V(\pi)$.  
(Indeed, we can show \cite{empiricalMnu08}
$U_{ed}= U_{ue}^\dagger U_{ud}=V^\dagger(\delta_{ue}) V(\delta)
 = {\bf 1} +{\cal O}(|V_{ub}|)$, so that we can still 
consider $U_{ue} \simeq U_{ud}$).
Therefore, the case $U_{ue}=V(\pi)$ is likely.
However, in this paper, we a priori assume the form
$U_{ue}=V(\delta_{ue})$ with $\delta_{ue} \geq 2\pi/3$ 
as a phenomenological requirement suggested in Table 1. 
Again, we summarize our phenomenological neutrino mass 
matrix which can lead to a nearly tribimaximal mixing for 
$|\delta_{ue}|\geq 2\pi/3$ as follows:
$$
(M_\nu)_e = k_\nu Y_e^D \left[ \left(
 V^T(\delta_{ue})\Phi_u^D V(\delta_{ue})\right)
 Y_e^D
+ Y_e^D  \left(V^T(\delta_{ue})\Phi_u^D V(\delta_{ue})
\right) \right]^{-1} Y_e^D  ,
\eqno(2.17)
$$
where 
$Y_e^D \propto {\rm diag}(m_e, m_\mu, m_\tau)$ and
$\Phi_u^D \propto {\rm diag}(\sqrt{m_u}, \sqrt{m_c}, 
\sqrt{m_t})$.
(For the phenomenological reason why the mass matrix 
(2.17) can give a nearly tribimaximal mixing, see 
Ref.\cite{empiricalMnu08}.)


As seen in Table 1, 
the predicted value of $R=\Delta m^2_{21}/\Delta m^2_{32}$
is considerably small compared to the observed value
$|R|=0.028\pm 0.004$,
where we have used the observed values 
$\Delta m^2_{21}=(7.59\pm 0.21) \times 10^{-5}\, {\rm eV}^2$
\cite{KamLAND} and 
$|\Delta m^2_{32}|=(2.74^{+0.44}_{-0.26}) \times 10^{-3}\, {\rm eV}^2$
\cite{MINOS}.
The value $R$ can be adjusted by taking the $\xi_0$-term in 
Eq.(1.4) into consideration.  
(It is easy to bring the $\xi_0$-term into the present model.)  
However, the smallness of $\Delta m^2_{21}$ can also become
mild by considering the renormalization group equation (RGE) effects.
Since, so far, we have not fixed the energy scale $\Lambda$,  
the values without RGE effects have been listed in Table 1. 
The RGE effects will be able to give a reasonable value
of $R$ without the $\xi_0$ term.
By the way, the present neutrino masses are normal hierarchical, 
so that, if 
regard $m_{\nu 3}$ as $m_{\nu 3}=\sqrt{\Delta m^2_{32}}=0.0523$
eV, we can obtain neutrino masses
$m_{\nu 1}=0.78$ meV, 
$m_{\nu 2}=1.52$ meV
and $m_{\nu 3}=52.3$ meV for the case $\delta_{ue}=\pi$.  

\vspace{3mm}

{\large\bf 3 \ Down-quark sector}

So far, we did not discuss a structure of $W_d$.
Although the purpose of the present paper is not to
give a structure of $W_d$, here, let us discuss 
a possible structure of $W_d$ lightly.
As we have assumed that the fields $\Phi_u$ and $\Phi_e$
are real and since we know that the $CP$ is broken 
in the quark sector, we must consider that $Y_d$ is
complex.
By way of trial, let us assume the following superpotential 
$W_d$:
$$ 
W_d =\lambda_{du} \left( {\rm Tr}[\Phi_X] {\rm Tr}[\Phi_u X_d]
+e^{i\alpha}{\rm Tr}[\Phi_u]  {\rm Tr}[\Phi_X X_d] \right)
+ m_d {\rm Tr}[Y_d X_d] + \lambda_d \, {\rm det}\Phi_{X} ,
\eqno(3.1)
$$
where we have assumed U(1)$_X$ charges $Q_X(X_d) = -Q_X(Y_d) 
\equiv -q_d$ and $Q_X(\Phi_X) =q_d -\frac{1}{2} q_u$. 
Since we consider $\Phi_u$ and $\Phi_X$ are real, the factor 
$e^{i\alpha}$ in Eq.(3.1) has been introduced by hand 
in order to yield a $CP$ violating
phase. 
Under this charge assignment, the term 
${\rm Tr}[\Phi_X \Phi_u X_d]$ is also allowed.
So far, in $W_u$, $W_e$ and $W_R$, we have not considered
cubic terms of a type ${\rm Tr}[A] {\rm Tr}[BC]$, while, 
in $W_d$, we have assumed such a cubic term 
${\rm Tr}[A] {\rm Tr}[BC]$ instead of a cubic term
${\rm Tr}[ABC]$.
At present, the form (3.1) is merely a phenomenological
assumption, and the form (3.1) is not a general form.
Also note that the cubic term ${\rm det}\Phi_{X}$ breaks the
U(1)$_X$ symmetry.
From the condition ${\partial W}/{\partial \Phi_{X}} = 0$, 
we obtain 
$$
Y_d = - \frac{\lambda_{du}}{m_d} ({\rm Tr}[\Phi_X] \Phi_u +
e^{i\alpha}{\rm Tr}[\Phi_u] \Phi_X ).
\eqno(3.2)
$$
Since we have already taken 
$\partial W_u/\partial \Phi_u =0$ in Eq.(2.4), 
we obtain $X_d=0$ for $\Phi_X \neq 0$ 
from the condition ${\partial W}/{\partial \Phi_{u}}
= \lambda_{du}({\rm Tr}[\Phi_X] X_d+ e^{i\alpha} 
{\rm Tr}[\Phi_X X_d]{\bf 1} ) +
\partial W_u/\partial \Phi_u = 0$. 
Then, from the condition ${\partial W}/{\partial \Phi_X} = 0$, 
we obtain 
$$
0= \frac{\partial {\rm det}\Phi_X}{\partial \Phi_X} 
= \Phi_X \Phi_X -{\rm Tr}[\Phi_X] \Phi_X
+({1}/{2})\left( {\rm Tr}^2[\Phi_X]-{\rm Tr}[\Phi_X\Phi_X]
\right) {\bf 1} ,
\eqno(3.3)
$$
where we have used a formula for a
$3\times 3$ Hermitian matrix $A$, 
${\rm det}A = ({1}/{3}) {\rm Tr}[AAA] 
-({1}/{2}) {\rm Tr}[AA] {\rm Tr}[A] 
+({1}/{6}) {\rm Tr}^3[A]$.
The constraint (3.3) demands 
that the matrix $\langle\Phi_X\rangle$ is a rank-1 matrix.
Such a rank-1 matrix is generally expressed as
$(\langle\Phi_X\rangle_u)_{ij} = v_X x_i x_j$,
where $x_i$ are real and $x_1^2+x_2^2+x_3^2=1$. 
Therefore, $\langle Y_d\rangle_u$ is expressed as
$$
(\langle Y_d\rangle_u)_{ij} \propto 
 \delta_{ij}  \sqrt{m_{u i}} 
+e^{i\alpha}x_i x_j (\sqrt{m_{u 3}}+\sqrt{m_{u 2}}
+\sqrt{m_{u 1}})  ,
\eqno(3.4)
$$
so that we obtain
$$
\frac{m_{s}}{m_{b}} \simeq 
\frac{1}{2} \sqrt{
\frac{m_{c}}{m_{t}} } ,
\eqno(3.5)
$$
where we have assumed $e^{i\alpha}\simeq 1$ and 
$(x_2/x_3)^2 \ll \sqrt{u_{u2}/m_{u3}}$.
The observed values are $m_{s}/m_{b} \simeq 0.019$
and $\sqrt{{m_{c}}/{m_{t}}} \simeq 0.060$ at $\mu=M_Z$
\cite{FK-qmass}, so that the relation (3.5) is in roughly
agreement with the observed value.
(We can adjust the predicted value to the observed value
by taking a suitable choice of $x_i$ and $\alpha$.) 
Also we can obtain $m_d/m_s \simeq \sqrt{{m_{u}}/{m_{c}}}$,
but the result is sensitive to the values of $x_i/x_j$
and $\alpha$, so that we do not discuss no more details 
of $m_{di}/m_{dj}$ in this paper.


\vspace{3mm}

{\large\bf 4 \ Concluding remarks}

In conclusion, we have proposed a new approach to the masses and
mixings of quarks and leptons. 
In the new approach, we write a superpotential $W$ for 
O(3)-flavor ${\bf 1}+{\bf 5}$ fields $Y_f$ whose VEVs give 
effective Yukawa coupling
constants $\langle (Y_f)_{ij}\rangle/\Lambda$
and we obtain relations among masses and mixings from
the SUSY vacuum conditions.
In this approach, we cannot predict the absolute values of
the masses and mixings, but we can obtain relations among
the VEV matrices $Y_f$ and $Y_R$.
Under this approach, we have found an empirical neutrino mass
matrix (2.17).
The form (2.17) was found as a byproduct when we assumed that
$\langle Y_e\rangle$ and $\langle Y_d\rangle$ can 
simultaneously be diagonalized.
Regrettably, the idea $U_{ue}=U_{ud}=V(\delta)$ with 
$\delta\simeq \pi/3$ was  failed to explain the observed
fact $\sin^2 2\theta_{23} \simeq 1$, but 
we have found that $U_{ue}=V(\delta_{ue})$ with
$\delta_{ue} \geq 2\pi/3$ can successfully give the 
observed values $\sin^2 2\theta_{23} \simeq 1$ and
$\tan^2\theta_{12}\simeq 0.5$.
At present, there is no theoretical reason for the form 
$U_{ue}=V(\delta_{ue})$. 
(Since we have assumed that $\Phi_u$ and $\Phi_e$ are real,
$U_{ue}$ must be real.) 
Nevertheless, it is worthwhile noticing because the form is
of a new type which is related to the up-quark masses and
which successfully leads to the nearly tribimaximal mixing
without assuming any discrete symmetry. 
(However, we do not consider that this denies applicability
of a discrete symmetry to the neutrino sector.
Rather, we consider that this suggests that
the discrete symmetry is applicable not only to the lepton
sector, but also to the quark sector.)

Since the present approach is still in its beginning stage,
we have many tasks to investigate: for example, 
(i) investigation of the explicit structures of 
$W_{\Phi u}(\Phi_u)$ and $W_{\Phi e}(\Phi_e)$, which 
completely determine the eigenvalues of $\langle\Phi_u\rangle$
and $\langle\Phi_e\rangle$, i.e.
$(\sqrt{m_u}, \sqrt{m_c}, \sqrt{m_t})$ and 
$(\sqrt{m_e}, \sqrt{m_\mu}, \sqrt{m_\tau})$, respectively; 
(ii) investigation of the explicit structure of $W_d$ 
in order to give more definite quark mass relations and
CKM matrix parameters ($Y_d$ in this paper 
has included free parameters $x_i$, so that we cannot 
derive definite conclusions because  
we can adjust the parameters $x_i$ to 
the observed values freely);
(iii) investigation of symmetries and quantum number 
assignments which can uniquely derive the present specific 
(phenomenological) form of $W$. 
In the present scenario, most of the fields $\Phi_u$, $\Phi_e$,
$Y_f$ ($f=u,d,e$), $Y_R$, and so on, take VEV of the
order of $\Lambda$, and their masses are also of the order $\Lambda$.
However, some components of those fields are massless in the
SUSY limit, and, under the SUSY breaking at $\mu\sim 1$ TeV,
they have masses of the order $\mu \sim 1$ TeV.
Since those particles are gauge singlets, in principle, they
are harmless in the low energy phenomenology.
However, in TeV region physics, we may expect fruitful 
phenomenology about flavor-mediated (but gauge-singlet) processes.
This approach will shed a new light on the quark and lepton
masses and mixings and on a TeV scale flavor physics.

\vspace{6mm}

\centerline{\large\bf Acknowledgments} 
The author would like to thank T.~Yamashita for 
helpful discussions, especially on the O(3) symmetry.
This work is supported by the Grant-in-Aid for
Scientific Research, Ministry of Education, Science and 
Culture, Japan (No.18540284).




\end{document}